\documentclass[notitlepage,twocolumn,superscriptaddress,amsmath,amssymb,aps,prx]{revtex4-2}
\pdfoutput=1
\usepackage[utf8]{inputenc}
\usepackage[T1]{fontenc}
\usepackage{hyperref}
\usepackage{graphicx}
\usepackage{amsfonts, amsmath, amsthm, amssymb} 
\usepackage{braket}
\usepackage{float}
\usepackage{amsthm}

\usepackage{color}
\usepackage[all]{nowidow}
\usepackage[dvipsnames]{xcolor} 
\hypersetup{citecolor=black,colorlinks=false,urlcolor=black} 
\usepackage[english]{babel}

\newcommand{\tr}[1]{\text{tr}\left(#1\right)}
\newcommand{\lindblad}[1]{\mathcal{L}\left(#1\right)}

\usepackage{bbold}
\usepackage{csquotes}
\usepackage{mathrsfs}
\begin{document}

\title{Bounding many-body properties under partial information\\ and finite measurement statistics
}

\author{Luke Mortimer}
\affiliation{ICFO - Institut de Ciencies Fotoniques, The Barcelona Institute of Science and Technology, 08860 Castelldefels, Barcelona, Spain}
\affiliation{Qilimanjaro Quantum Tech., Carrer de Venecuela, 74, Sant Marti, 08019 Barcelona, Spain}

\author{Leonardo Zambrano}
\affiliation{ICFO - Institut de Ciencies Fotoniques, The Barcelona Institute of Science and Technology, 08860 Castelldefels, Barcelona, Spain}

\author{Antonio Ac\'{\i}n}
\affiliation{ICFO - Institut de Ciencies Fotoniques, The Barcelona Institute of Science and Technology, 08860 Castelldefels, Barcelona, Spain}
\affiliation{ICREA - Instituci\'{o} Catalana de Recerca i Estudis Avan\c{c}ats, 08010 Barcelona, Spain.}

\author{Donato Farina}
\affiliation{Physics Department E. Pancini, Università degli Studi di Napoli Federico II, Complesso Universitario Monte S. Angelo, Via Cinthia, 21, Napoli, 80126, Italy}
\affiliation{Istituto Nazionale di Fisica Nucleare (INFN), Sezione di Napoli, Complesso Universitario Monte S. Angelo, Via Cinthia, 21, Napoli, 80126, Italy.
}

\date{\today}

\begin{abstract}
Calculating bounds of properties of many-body quantum systems is of paramount importance, since they guide our understanding of emergent quantum phenomena and complement the insights obtained from estimation methods. Recent semidefinite programming approaches enable probabilistic bounds from finite-shot measurements of easily accessible, yet informationally incomplete, observables. Here we render these methods scalable in the number of qubits by instead utilizing moment-matrix relaxations. After introducing the general formalism, we show how the approach can be adapted with specific knowledge of the system, such as it being the ground state of a given Hamiltonian, possessing specific symmetries or being the steady state of a given Lindbladian. Our approach defines a scalable real-world certification scheme leveraging semidefinite programming relaxations and experimental estimations which, unavoidably, contain shot noise.
\end{abstract}

\maketitle

\section{Introduction}

{
The study of many-body systems remains a central theme in contemporary physics\,\cite{RevModPhys.91.021001, amico2008entanglement}.  
Many-body interactions give rise to emergent collective behaviours such as phase transitions, and lead to paradigmatic phenomena including superradiance\,\cite{gross1982superradiance, RevModPhys.91.035003, masson2022universality}, superfluidity and superconductivity\,\cite{RevModPhys.29.205, RevModPhys.78.17}, and topological phases of matter\,\cite{RevModPhys.89.041004, PhysRevX.8.031079}. 
Beyond their fundamental interest, understanding and characterizing many-body quantum systems is also crucial both for validating experiments, and for the development of modern quantum devices, where quantum noise limits their scalability and thus their potential quantum advantages. This includes advances in quantum computing, quantum materials, quantum simulators, and quantum information science in general \cite{amico2008entanglement, preskill2018quantum}.
Consistent with their potential quantum advantages, exact classical simulation of many-body systems is generally intractable due to exponential scaling in the number of constituents \cite{Feynman1982Simulating}. On the experimental side, full quantum state tomography requires prohibitively many measurements \cite{haah2017SampleOptimal}.
Classically computing system properties 
becomes even more demanding in realistic open-system settings, where states are represented by density matrices \cite{fazio2025many}.

This calls for scalable methods that can either provide estimates of many-body properties or rigorous bounds. Quantities of interest may be linear observables, such as local magnetization or correlation functions, or nonlinear functions such as purity, fidelity, or entanglement measures.
On one hand, estimation methods based on a variational ansatz, such as tensor networks \cite{RevModPhys.93.045003} or quantum Monte Carlo \cite{foulkes2001quantum}, are established tools in condensed matter physics. They find estimates of physical quantities, but they typically are not able to provide rigorous bounds. On the other hand, convex optimization offers a natural and scalable framework for certification tasks. 
In many-body physics, such techniques are less established than variational methods, thereby opening the way for new approaches and applications.
A convex optimization problem consists of minimizing a convex function over a convex set. Such problems guarantee convergence to the global optimum (i.e., there are no issues with local optima). Consequently, they enable the derivation of rigorous, inviolable bounds on physical quantities.
In particular, semidefinite programming (SDP) is a subclass of convex optimization in which the objective is a linear function. It is a well-established tool
in computer science and engineering \cite{boyd2004convex, vandenberghe1996semidefinite} as well as in quantum information theory \cite{skrzypczyk2023semidefinite}.
SDP approaches are made scalable through SDP relaxations (SDPR), which have a long history in quantum information.
For instance, they enable bounding the set of quantum correlations \cite{Navascues2007NPA, Navascues2008NPA}, thereby
providing upper bounds on maximal Bell inequality violations.
Such bounds have a foundational relevance: measuring something beyond any of them would disprove the validity of quantum mechanics.

Notably, there is a growing interest toward applying convex optimization tools in many-body physics. 
SDPRs provide lower bounds on ground state energies.
This key result was first established in Ref.~\cite{pironio2010convergent} and subsequently found applications and further developments~\cite{baumgratz2012lower, PhysRevX.14.021008, PhysRevLett.108.263002}.
In condensed matter and quantum chemistry, a lower bound on the ground state energy complements upper bounds obtainable through variational methods. The two combined allow one to sandwich the true value.
This, when used as a constraint, can serve to bound other
ground state properties \cite{PhysRevX.14.031006}. 
SDPR find spectral gap certificates \cite{rai2024hierarchy} relevant for quantum annealing protocols.
In open quantum systems, leveraging convex optimization, one can correct non-Markovian master equations that violate physical requirements \cite{PhysRevX.14.031010}.
In many-body Markovian systems, SDPRs can also be used to obtain Liouvillian gap certificates that provide lower bounds on the slowest decay rate \cite{chen2024boosting}, which is relevant for certifying fast thermalization. 
They find an upper bound on the fastest decay rate \cite{mok2024universal}, fundamental to identify timescales where the system is, instead, fairly insensitive to environmental disturbance. Furthermore, SDPRs can bound thermal state features \cite{fawzi2024certified}, and generic steady-state properties of Lindblad dynamics in finite dimensional (e.g., many spins) \cite{us} and bosonic  \cite{robichon2024bootstrapping} systems. In quantum thermodynamics, different SDPR enable upper bounds on the maximum extractable work from open quantum batteries
\cite{PhysRevA.107.012405, PhysRevA.111.012212}. 
Additionally, SDPR have been proposed as tools to witness phase transitions in closed \cite{jansen2025mapping} and open \cite{cho2025nonequilibrium}-system scenarios.
Hence, semidefinite programming techniques find relevant applications in the characterization of both closed and open many-body quantum systems.

On the experimental side, reconstruction methods use measurement data from the actual system to estimate expectation values and determine confidence regions \cite{guctua2020fast, gross2010CompressedSensing, huang2020predicting, zambrano2024certification}. 
While SDPR methods can exploit physical constraints but ignore measurement data, reconstruction methods incorporate experimental statistics but do not fully leverage prior physical knowledge of the system.
In other words, certification schemes under partial information and shot noise are, in their current form \cite{zambrano2024certification}, nonscalable in the number of constituents.
Nonetheless, they represent a step forward in bridging SDP methods and empirical information, deriving probabilistic bounds.  
}

{In this work, we introduce a scalable real-world certification scheme leveraging semidefinite programming relaxations and experimental estimations.
This hybrid approach allows us to compute rigorous bounds for various system properties, including paradigmatic expectation values of observables and nonlinear functions such as state purity. These bounds are  tighter than those obtained using either simulations or measurements alone.
We show this visually in Figure \ref{fig:bounds}.
Before proceeding, we emphasize that for a completely general many-body property, the number of variables required to characterize the problem may scale exponentially with system size. As such, no polynomial-time certification procedure can be expected in full generality. In this work, we therefore focus on physically motivated scenarios in which the relevant set of variables (concretely, for this work, the moments that we will introduce in Sec.\,\ref{sec:relaxation}) grows at most polynomially, and for which the resulting SDP relaxations remain tractable. Many problems of practical interest---such as those involving local Hamiltonians, certain open systems, and few-body observables---naturally satisfy this requirement.

{
The article is organized as follows. 
In Sec.\,\ref{sec:formalism} we introduce our general tool to derive rigorous probabilistic bounds on many-body system properties leveraging SDP relaxations and concentration inequalities on measurement estimations.
In Sec.\,\ref{sec:numerical-results} we benchmark our method on a variety of many-spin models and using simulated measurements. They include both open and closed dynamics as well as large systems, with the goal of bounding both linear and non-linear functions of the quantum state. 
In particular, in the open scenario, we analyse a two-dimensional grid of qubits with nearest-neighbour interactions and transverse field, coupled on opposite edges to hot and cold thermal baths. This model reaches a non-equilibrium steady state, known to be hard to identify and serves as a realistic benchmark for the technique. 
We will show that, even when only a small subset of observables is measured, incorporating partial measurement information into the SDP relaxation leads to increased accuracy. This is  particularly evident for observables involving complex multi-qubit Pauli strings that are costly to measure directly.
Altogether, these features will demonstrate the wide applicability and the potential of the method.
In Sec.\,\ref{sec:conc}, we finally summarize our conclusions and we draw perspectives for future developments.
}

\section{General formalism}
\label{sec:formalism}

%

\begin{figure}
    \centering
    \includegraphics[width=0.99\linewidth]{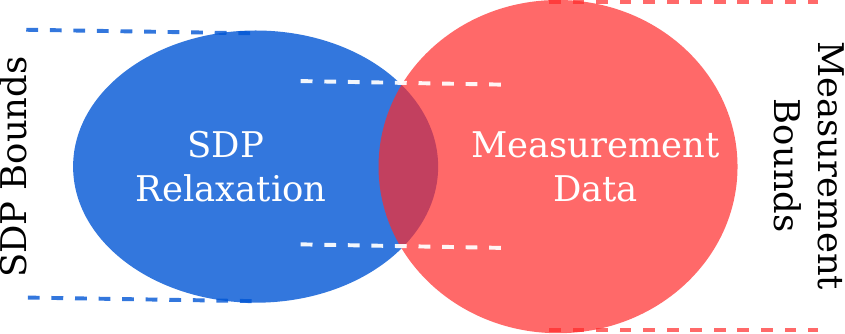}
    \caption{A visualization demonstrating the comparison between bounds obtained from an SDP relaxation, from measurement data, and from the combined set. Note how the combined set 
    has tighter bounds than either set individually. }
    \label{fig:bounds}
\end{figure}

\subsection{Bounds on expectation value estimates}

In this work, we restrict our attention to finite-dimensional systems.
Following Refs.\,\cite{PhysRevLett.124.100401, zambrano2024certification},
let $M_{i}$ be an observable with eigenvalues in the interval $[-1, 1]$.
This entails no loss of generality, since any observable can be linearly transformed to have spectrum contained in this interval by an appropriate shift and rescaling.
Repeated $N_i$ projective  measurements of the observable $M_{i}$ over a quantum state $\rho$ define independent, identically distributed, random variables $X_i^{(j)} \in [-1,1]$ for $j=1,2, \dots, N_i$. The expectation value of $X_i^{(j)}$ is then given by $\mathbb{E} \left[X_i^{(j)}\right] = \mathrm{tr}(M_{i} \rho)$, and the empirical mean over $N_i$ samples by
\begin{equation}
\bar{m}_i = \frac{1}{N_i}\sum_{j=1}^{N_i} X_i^{(j)}.
\end{equation}
To bound the probability that $\bar{m}_i$ deviates from $\mathrm{tr}(M_{i} \rho)$, we apply Hoeffding’s inequality \cite{hoeffding1963probability} for bounded independent variables:
\begin{equation}
\Pr\left(|\bar{m}_i - \mathrm{tr}(M_{i} \rho)| \geq \epsilon_i\right) \le 2 \exp\left(-\frac{2 N_i \epsilon_i^2}{\sigma^2}\right),
\end{equation}
where $\sigma = \max_j \left|X_i^{(j)} - \mathrm{tr}(M_{i} \rho)\right|\leq 2$. Substituting $\sigma=2$ yields the simplified form
\begin{equation}
\Pr\left(|\bar{m}_i-\mathrm{tr}(M_{i} \rho)| \geq  \epsilon_i\right) \le 2\exp\left(-\frac{N_i\epsilon_i^2}{2}\right).
\end{equation}
This bound controls the probability that the estimate $\bar{m}_i$ deviates by more than $\epsilon_i$ from $\mathrm{tr}(M_{i} \rho)$. 

If $K$ observables $M_1, \dots, M_K$ are measured (in parallel or sequentially), 
and we wish to ensure that \textit{all} empirical means simultaneously satisfy their own deviation bound, we apply the union bound \cite{boole1847mathematical}. 
Specifically, the probability that \textit{any} of the $K$ estimates violates the bound is at most the sum of their individual failure probabilities. Let $\delta$ be the desired overall failure probability. Setting $2 \exp\left(-{N_i \epsilon_i^2}/{2}\right) = {\delta}/{K}$ ensures that the probability of \textit{any} deviation exceeding $\epsilon_i$ is bounded by $\delta$. 
Solving for $\epsilon_i$ 
gives, 
%
\begin{equation}
\epsilon_i(\delta) = \sqrt{\frac{2\log(2K/\delta)}{N_i}}\,.
\label{eqn:epsilon}
\end{equation}
Therefore, the probability that $|\langle M_{i} \rangle - \bar{m}_i| \leq \epsilon_i$
for all $i$ is at least $1-\delta$, with $\epsilon_i$ decreasing as the number of shots increases. These constraints form a convex set and can therefore be incorporated into an SDP (as shown in Ref.\,\cite{zambrano2024certification}).

\subsection{Exact SDP problem}
Considering a quite general case,
we are interested in lower bounding the expectation value of a given observable $\mathcal{O}$ over all quantum states compatibles with some constraints.
Such constraints may include both measurement data and additional guarantees.
This can be cast as an optimization problem, given by
\begin{eqnarray}
\label{exact-sdp}
{o}_{\rm min} \,=\,\min_{\rho}\quad && \tr{\mathcal{O}\rho} \\
    \text{  s.t.} \quad &&  \rho \succcurlyeq 0\nonumber\,,\, \tr{\rho}=1\nonumber\\
   && \vert \tr{\rho M_i}-\bar{m}_i \vert \leq 
   \epsilon_i(\delta)
   \quad \forall \, i \in [K]
   \nonumber\\
   && \ell^{(j)}(\rho)=0 \nonumber \quad \forall \, j \in [S]\,.
\end{eqnarray}
The bound holds true with a probability at least $1-\delta$, as a consequence of measurement confidence regions.
This turns out to be the formalism of Ref.\,\cite{zambrano2024certification}, 
with additional linear constraints on the state variables (last constraint in \eqref{exact-sdp}, $\ell$ is a linear superoperator). We suppose to have $S$ of these constraints. They may, for instance, represent assumed symmetries or that the state be a \enquote{zero} of a Lindbladian, i.e. a steady state, with more explicit examples illustrated later.
Notice that by instead minimizing $\tr{  \rho \mathcal{O'}}$, with $\mathcal{O'}=-\mathcal{O}$, we would have obtained $o'_{\min}=-o_{\rm \max}$, allowing one to eventually identify an interval 
$[o_{\min}, o_{\max}]$ for the 
value of $\mathcal{O}$ on the state of interest.

Problem \eqref{exact-sdp} is a semidefinite program. As such, it can be solved efficiently in the problem's number of variables and constraints finding a global optimum.
However, let us consider a system of $n$ qubits. The number of variables of the problem, the $4^n-1$ degrees of freedom of $\rho$, scales exponentially in the system size $n$. This, in turn, makes the computation time scale exponentially in $n$, as $poly(expo(n))=expo(n)$.
\subsection{Relaxed problem}
\label{sec:relaxation}
From the previous discussion, $o_{\min}$ is typically inaccessible in large systems.
However, semidefinite programming relaxations  allow one to keep the execution time 
$poly(n)$ whilst generally paying the price of obtaining looser bounds.
SDPR will return a value, being a lower bound (lb) for $o_{\min}$, such that $o_{\rm lb}\leq o_{\min}$.
Following the previous reasoning, the method also allows one to obtain an upper bound, eventually identifying an interval 
\begin{equation}
\label{relation-with-exact}
    [o_{\rm lb}, o_{\rm ub}] \supseteq [o_{\rm min}, o_{\rm max}]\,.
\end{equation}
We show how to obtain this relaxation step by step, in line with the methods used in the Navascués–Pironio–Acín (NPA) hierarchy \cite{Navascues2007NPA, Navascues2008NPA}.
\subsubsection{Change of variables: moments}
First of all, we perform a change of variables. 
Instead of the entries of $\rho$, we consider as new problem variables the \textit{moments} $\langle P_\alpha \rangle=\tr{\rho P_\alpha}$, with the $P_\alpha$'s being suitable operators for the problem at hand.
We suppose our problem can be described by a number $\vert \mathcal{I}_v \vert $, cardinality of the set $\mathcal{I}_v$, of variables $\{ \langle P_\alpha \rangle \}_{\alpha  \in \mathcal{I}_v}$.
For simplicity, we assume the operators $\{ \langle P_\alpha \rangle \}_{\alpha  \in \mathcal{I}_v}$ are Hermitian and their expectation value is bounded between $-1$ and $1$.  We also require they form a subset of an operator basis $\{ \langle P_\alpha \rangle \}_{\alpha=1}^{d^2}$, $d$ being the Hilbert space dimension.
\subsubsection{Objective function and observables}
We can conveniently represent the objective function in this basis,
$\tr{\rho \mathcal{O}}=\sum_{\alpha \in \mathcal{I}_o} o_{\alpha} \braket{P_{\alpha}}$. 
The real coefficients $o_{\alpha}$ are assumed to have support on a set $\mathcal{I}_o \subseteq \mathcal{I}_v$.
Analogously for the measured observables $\{M_k\}_k$ we have
$\tr{\rho M_k}=\sum_{\alpha \in \mathcal{I}_{M_k}} m_{k,\alpha} \langle {P_{\alpha}} \rangle$.

\subsubsection{Relaxing the positivity constraint}
The positivity and trace-one condition of the state imply the bounds $-1\leq \braket{P_{\alpha}} \leq 1$. 
Less trivially, they imply the positivity of moment matrices and reduced density matrices.
Imposing moment-matrix positivity amounts to requiring the 
positive semi-definiteness of a square matrix $\mathcal{M}_k$, of components 
$(\mathcal{M}_{k})_{\alpha, \beta}=\tr{P_\alpha P_\beta \rho}$, where $\alpha$ index a scalable subset of basis operators.
Indeed, $\rho \succcurlyeq 0$ implies  $\mathcal{M}_k \succcurlyeq 0$, but, in general, not vice-versa.
This property qualifies the approach as a relaxation.
By exploiting algebraic properties (such as Pauli reduction rules when considering spin systems),
one obtains 
$\mathcal{M}_{k}=B_k+\sum_{\alpha \in \mathcal{I}_{+_k}} A_{k,\alpha} \braket{P_{\alpha}} \succcurlyeq 0$, where the $B_k$'s and the $A_{k,\alpha}$'s are appropriate matrices (by construction, $\mathcal{I}_{+_k} \subseteq \mathcal{I}_v$ for all $k$). 
This expression renders explicit the linear combination of moments with given matrix coefficients.
The positivity constraints of reduced density matrices (RDMs) take analogous forms. In Appendix~\ref{app:pos}, we provide explicit, pedagogical examples illustrating the positivity of the moment matrix and of reduced density matrices. 
Here we suppose to impose a number $k_+$ of such positivity conditions.
%
%

\subsubsection{Symmetries and other guarantees}
Finally, each given symmetry or guarantee, i.e. each linear constraint $j \in [S]$ on the state $\rho$ appearing in \eqref{exact-sdp}, implies in general a number $\kappa_j$ of linear of constraints on the moments. 
\subsubsection{General scalable problem}
Leveraging these considerations, the general relaxed problem reads as follows:
\begin{align}\label{sdp-notation}
o_{\rm lb}=  & \min_{ \Huge{\{\braket{P_{\alpha}}  \, : \,  \alpha \in \mathcal{I}_v \}}}   \quad  \textstyle{\sum_{\alpha \in \mathcal{I}_o}} o_{\alpha} \braket{P_{\alpha}} \\
 & \text{s.t. } \;  -1\leq \braket{P_{\alpha}} \leq 1 \qquad \qquad\qquad\quad\;  \forall\, \alpha\in \mathcal{I}_v \nonumber \\
& \qquad B_k+\textstyle{\sum_{\alpha \in \mathcal{I}_{+_k}}} A_{k,\alpha} \braket{P_{\alpha}} \succcurlyeq 0 \qquad  \: \; \;  \forall \,k \in [k_+]   \nonumber \\
& \qquad  \big\vert\textstyle{\sum_{ \alpha \in \mathcal{I}_{M_i}}} m_{i,\alpha} \braket{P_{\alpha}}-\Bar{m}_i \big\vert \leq \epsilon_i(\delta)\quad \forall \; i \in [K]  
\nonumber \\
& \qquad  \textstyle{\sum_{\alpha \in \mathcal{I}_{S_{j, k}}}} l_{j, k,\alpha} \braket{P_{\alpha}} = 0 
\quad 
\forall \, j \in [S]\,,
\, 
\forall \, k \in [\kappa_j]
\,. \nonumber
\end{align}

To guarantee \textit{scalability} in the number $n$ of constituents, both 
the number of constraints
and the number of variables should scale polynomially with $n$. 
In our notation,
the numbers $k_+$, $K$, $S$, $\sum_{j=1}^S\kappa_j$ and $\vert \mathcal{I}_v\vert$ must grow polynomially in $n$. Notice that the set $\mathcal{I}_v$, identifying the variables of the problem \eqref{sdp-notation}, is the union of all the sets of interest in the problem.
In formulas, we have   
$
\mathcal{I}_v = \mathcal{I}_o
\cup \mathcal{I}_+
\cup \mathcal{I}_M
\cup \mathcal{I}_S
$,
with
$\mathcal{I}_M:=\cup_{k \in [K]} \mathcal{I}_{M_k}$, 
$\mathcal{I}_+:=\cup_{k \in [k_+]} \mathcal{I}_{+_k}$
and
$\mathcal{I}_S:=\cup_{j \in [S]}  \cup_{k \in [\kappa_j]} \mathcal{I}_{S_{j, k}}$.
A summary of the index sets used to label the moment
variables of interest is reported in Table\,\ref{tab:sets}.
At variance with problem \eqref{exact-sdp},
the construction in Eq.\,\eqref{sdp-notation} ensures that the computation time now scales polynomially with the system size, since $poly(poly(n))=poly(n)$.
\begin{table}[t]
\label{tab:sets}
\caption{Summary of the index sets used to label the moment variables of interest.}
\label{tab:sets}
\vspace{.2cm}

\centering
\begin{tabular}{|c|p{3.7cm}|p{2.7cm}| }
\hline
\textbf{Index set} & \textbf{Labelling moments in}& \textbf{Structure} \\
\hline\hline
$\mathcal{I}_v$& full problem \eqref{sdp-notation}& $\mathcal{I}_o \cup \mathcal{I}_+ \cup \mathcal{I}_M \cup \mathcal{I}_S $ \\ \hline 
$\mathcal{I}_o$& objective function&  \\ \hline
$\mathcal{I}_+$& moment matrices/RDMs& $\cup_{j\in [k_+]} {\mathcal{I}_{+}}_j$  \\ \hline
$\mathcal{I}_M$& {measured {observables}}&  $\cup_{j\in [K]}{\mathcal{I}_M}_j$\\ \hline
$\mathcal{I}_S$& symmetries/guarantees&  
$\cup_{j\in [S]}\cup_{k\in [\kappa_j]} \mathcal{I}_{S}{}_{j,k}$\\ 
\hline 
\end{tabular}
\end{table}

In summary, problem \eqref{sdp-notation} allows one to  efficiently obtain rigorous bounds, defining the interval
\begin{eqnarray}
\label{relax-bounds}
    &[o_{\rm lb}, o_{\rm ub}]\,,&\\
    &\text{holding true at least with probability $1-\delta$}\,.& \nonumber
\end{eqnarray}
Nevertheless, we are not guaranteed our bounds are sufficiently tight to provide useful information in practice. The following examples serve to positively elucidate this point, showing that the obtained bounds are indeed informative, and to demonstrate the
highly general applicability of the formalism. 
Notably, our method is also applicable to convex non-linear functions such as that of purity, albeit restricted to finding only a lower bound due to the non-linearity.

\section{Numerical Results}
\label{sec:numerical-results}
We therefore have a certification tool that combines information from measurement data--in the form of bounds on expectation values of suitably chosen observables--and fundamental guarantees on the system--for instance, it being the steady state of a given Lindbladian, the ground state of a given Hamiltonian, or possessing certain symmetries--in the form of a  SDP relaxation. This framework is aimed at producing certified bounds on quantities of interest in many-body systems.

We note that related ideas, aimed at finding estimations of such quantities, have recently been explored in other contexts, such as combining shadow tomography with $N$-representability conditions for fermionic systems \cite{avdic2024fewer}, or using relaxations of reduced density-matrix compatibility for qubit systems \cite{wang2025mitigating}.
By contrast, our approach provides rigorous bounds at a prescribed confidence level, and is therefore complementary to these estimation-based methods.

In the following, we first discuss technical aspects of our scheme that are specific to the numerical simulations performed.
In particular, we describe the selection of the moment variables, the construction of the moment matrix, and the procedure used to artificially generate measurement data. We then present concrete use cases.
Under partial information, we demonstrate our method by bounding the steady-state heat current in an open quantum system and by obtaining a lower bound on its ground-state energy. 
We further demonstrate the applicability of our method to nonlinear functions by deriving lower bounds on subsystem purity when the full system is in its ground state.
Finally, we consider a $50$-qubit Majumdar–Ghosh model\,\cite{majumdar1969next}
to highlight the scalability of the method.

{
\subsubsection{Moment selection protocol}
In the numerical simulations presented here, we focus on \(n\)-qubit systems and choose as moment variables those associated with multi-qubit Pauli strings. A Pauli string is defined as a tensor product of single-qubit Pauli operators acting on an \(n\)-qubit system,
\begin{equation}
P_\alpha =
\sigma_{\alpha_1} \otimes 
\sigma_{\alpha_2} \otimes
\cdots \otimes
\sigma_{\alpha_n},
\end{equation}
with \(\sigma_{\alpha_i} \in \{\openone, X, Y, Z\}\).

We emphasize that there is not a single way to choose the scalable set $\mathcal{I}_v$ identifying the Pauli strings for the problem \eqref{sdp-notation}, nor to select the subset used to construct the moment matrix. 
However, different choices lead to different bounds for the objective function, and one should therefore look for a choice that yields sufficiently tight bounds. 
This aspect is itself an ongoing research direction, and pushing it further is not the main scope of the present work. Any choice does provide valid bounds, and we are satisfied whenever, under a reasonable moment-selection strategy, these bounds are sufficiently informative.

We now outline the general procedure followed in this work; any slight variations are highlighted on a case-by-case basis in the examples presented in Secs.~\ref{sec:SS}--\ref{sec:conf-region-plot}.
The first observation is that the most relevant moments to include are those that appear in the objective function (set $\mathcal{I}_o$). In fact, by construction, these moments must be included in the problem. 
The corresponding Pauli strings are then used to form linear equality constraints, which in turn typically generate additional Pauli strings (set $\mathcal{I}_S$). 
An example is the enforcement of the steady-state condition (see \cite{us} and Eq.\,\eqref{adj-lind-mom}) or the imposition of symmetries, depending on the problem at hand. 
This in general generates new moments for the problem.
We then include the moments associated to the measured observables (sets $\mathcal{I}_{M_k}$), which encode information obtained in the laboratory and enter the formulation as linear scalar inequalities.
Although more general choices are possible, in the following examples we consider a single moment matrix ($k_+=1$).
We identify the most common 
$\mathcal{N}$ moments appearing among the previously defined sets, and we use the associated 
$\mathcal{N}$ Pauli strings to build the matrix. This guarantees a positivity relaxation that is effective in tightening the bounds on the objective. 
While larger 
$\mathcal{N}$ values yield tighter or unchanged bounds, they also increase the computation time. For this reason, our simulations necessarily involve a compromise between tightness and tractability (this produces the set $\mathcal{I}_+$).

{\subsubsection{Measurement data generation}
In the following examples we artificially simulate measurement data. The chosen observables are directly a subset of Pauli strings $\{{P_\alpha}\}_{\alpha \in \mathcal{I}_M}$.
For simplicity, we assume that each operator $P_\alpha$, with $\alpha \in \mathcal{I}_M$, is measured with the same number of shots (i.e., independent copies of the state $\rho$), so that $N_i = N$ for all $i$.
The measurement outcomes are simulated using a binomial distribution determined by the probabilities $\{\mathrm{tr}(\rho P_\alpha^{+}), \mathrm{tr}(\rho P_\alpha^{-})\}$, where $P_\alpha^{+}$ and $P_\alpha^{-}$ denote the projectors onto the positive and negative eigenspaces of the Pauli operator $P_\alpha$, respectively. 
From these simulated outcomes, we compute the empirical sample mean $\bar{m}_\alpha$, which we use as an estimator of the corresponding expectation value \(\mathrm{tr}(\rho P_\alpha)\).
This procedure  allows us to obtain the constraints
$
\vert \langle P_\alpha \rangle 
-\bar{m}_\alpha \vert \leq 
\epsilon_\alpha(\delta)\,,
\, \forall \, \alpha \in \mathcal{I}_M
$.

We note that data generation performed in this manner is not scalable, as it requires working explicitly with the true density operator~$\rho$. This limitation is not an issue in practice, since the method~\eqref{sdp-notation} is intended to be supplied with real measurement data.
However, by considering appropriate classes of states, the artificial data generation can also be made scalable (see Sec.\,\ref{sec:large-system}), thereby allowing us to benchmark the scheme for large systems.
}

\subsection{Markovian open quantum systems: bounding steady-state properties}
\label{sec:SS}
We start by considering an open quantum system, whose dynamics is described by a suitable Lindblad generator.
The system is a
two-dimensional arrangement of qubits, with the leftmost qubits coupled to a hot bath at temperature $T_h$ and the rightmost qubits coupled to a cold bath at temperature $T_c$.
A pictorial representation is illustrated in Fig.\,\ref{fig:two-baths}.
The Lindbladian reads as follows,
\begin{equation}
  \lindblad{\rho}=-i\left[H, \rho\right] +\sum_{j}^n ( \gamma_j^{+} {D}[\sigma_{+}^{(j)}] \rho+\gamma_j^{-} D[\sigma_{-}^{(j)}] \rho ) \,, 
  \label{eqn:lindblad}
\end{equation}
with the Hamiltonian given by,
\begin{equation}
  H = g\sum_{i}^n Z_i + \frac{J}{2}  \sum_{\braket{i,j}}^n X_i X_j \,,
  \label{eqn:hamil}
\end{equation}
and $\langle i,j\rangle$ denoting nearest-neighbor pairs.
Here the superoperator $D$ is defined as  $D[A] B := A B A^\dagger -\frac{1}{2}\{A^\dagger A , B\}$ 
and the plus/minus jump operators as   $\sigma_{+}^{(j)}=\frac{1}{2}(X_j+iY_j)$ and $\sigma_{-}^{(j)}=\frac{1}{2}(X_j-iY_j)$. 
The kinetic coefficients $\gamma_j^\pm$ depend on the temperatures of the corresponding baths and on the site location.
For all the qubits on the left (right) edge, we have $\gamma_i^{\pm}=\gamma_{h(c)}^{\pm}$,
while for all the rest we have 
$\gamma_i^{\pm}=0$. 
Considering the left edge, they are given by
$ \gamma_h^{+}=\gamma_h n_{B, h}$
and
$\gamma_{h}^{-}=\gamma_{h} (n_{B, h}+1)$
where the Bose factor is $ n_{B,h} = 1 / (e^{\epsilon_h / T_h} - 1) $;
analogously for the right edge, replacing the label $h$ with $c$.
This is a dissipative version of the 2D transverse-field Ising model, whose steady state is driven out of thermal equilibrium by the imposed temperature bias.
This system was also analyzed in previous work~\cite{us}, but without leveraging measurement data.
\begin{figure}
    \centering
\includegraphics[width=.85\linewidth]{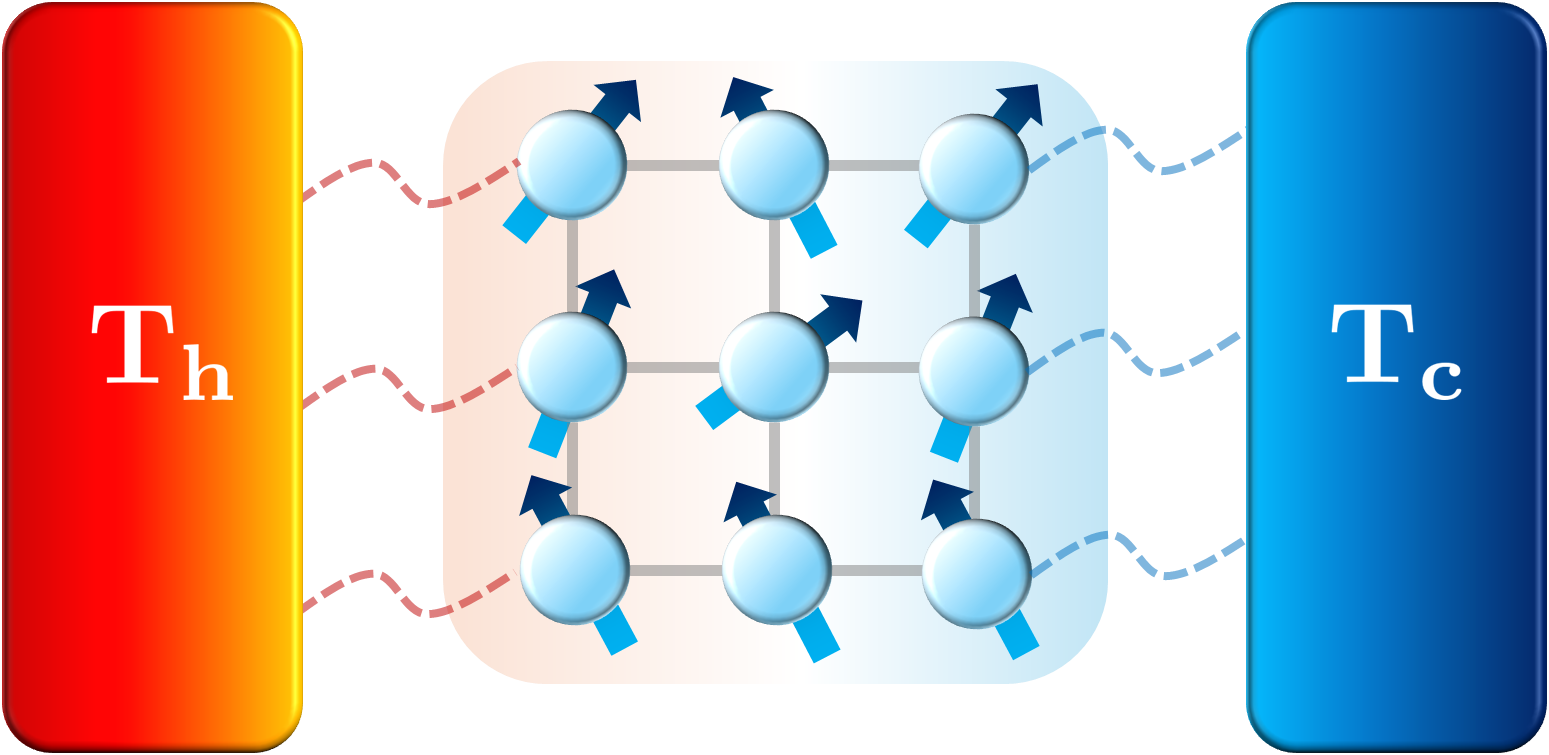}
    \caption{A pictorial representation of the open system described by Eq.\,\eqref{eqn:lindblad}.}
    \label{fig:two-baths}
\end{figure}
%
%

{Here, we are interested in bounding steady-state properties, incorporating such additional boost.
For a given Pauli string $P_\alpha$,
the steady state condition $\mathcal{L}(\rho)=0$ implies 
\begin{equation}
\label{adj-lind-mom}
\langle \mathcal{L}^\dag(P_\alpha)\rangle=0\,,
\end{equation}
where the adjoint Lindbladian $\mathcal{L}^\dag$ is defined implicitely by 
$
\langle A, \mathcal{L}(B) \rangle
=
\langle \mathcal{L}^\dagger(A), B \rangle
$ for all operators  
 $A,B$,
with 
$
\langle A, B \rangle \equiv {\rm tr}\left(A^\dagger B\right)
$
being the Hilbert–Schmidt inner product.
Computing the action of the adjoint Lindbladian over the Pauli string $P_\alpha$, one obtains a linear constraint in the form $\textstyle{\sum_{\alpha' \in \mathcal{I}_{S_{1, \alpha}}}} l_{1, \alpha,\alpha'} \braket{P_{\alpha'}} = 0$ of Eq.\,\eqref{sdp-notation}, 
having chosen the steady-state guarantee, here labelled as the guarantee $j=$ \enquote{$1$}.
This generally produces the new moment variables for the problem.
}

Concerning the objective function, the quantity we consider is the heat current
traversing the system, or equivalently
flowing out of the hot bath. 
This is given by 
\begin{equation}
    \langle \mathcal{O}\rangle  \equiv\langle \mathcal{L}_h(H) \rangle\,,
\end{equation}
where $\mathcal{L}_h$ is the dissipator associated to the effective interaction with the hot bath, and $H$ is the full system Hamiltonian.

We consider the case whereby $Z$ measurements on the qubits are not possible, perhaps because they are experimentally challenging or prone to errors. 
In this setting, the heat current cannot be measured directly, as it would involve $Z$ measurements. 
Instead, we only have access to the SDP relaxation and to measurements in the $X$ and $Y$ bases. Any information about $Z$ then comes indirectly, through the higher-order correlations captured by the SDP relaxation. 

The results are shown in Fig.~\ref{fig:heat}, where we present upper and lower bounds on the heat current as function of the the total number of shots $N_{\rm tot}$ used in the simulated experiment.
We report the bounds obtained using (blue curves) only measurement data, without additional constraints; (black curves) the full SDP formulation but without measurement data (as in Ref.~\cite{us}, the notion of shots is not meaningful here); and (red curves) the complete approach introduced in this work, which combines both sources of information.
Notably, the combined approach significantly outperforms the other two. 
For instance, among the three methods considered, this is the only one  able to certify a positive heat current, corresponding to transport from the hot to the cold bath, as expected from the second law of thermodynamics.

Regarding more technical aspects, the measurements were chose according to the following rule.
We chose the first (i.e. in order of generation) $100$ Pauli strings that appear in the Lindbladian and moment matrix constraints of our SDP and splitting the shots between only those. 
This means that the number of shots for each Pauli string considered is $N = N_{\mathrm{tot}} / 100$.
Many different strategies not shown in the figure were also tested.
They included measuring only those terms found in the objective, measuring all second- or third-order Pauli strings, measuring the Pauli strings which appear most frequently in our Lindbladian constraints, all with and without the SDP relaxation. However, none of them 
out-performed the corresponding methods of Fig.\,\ref{fig:heat}.

\begin{figure}
\centering
\includegraphics[width=0.99\linewidth]{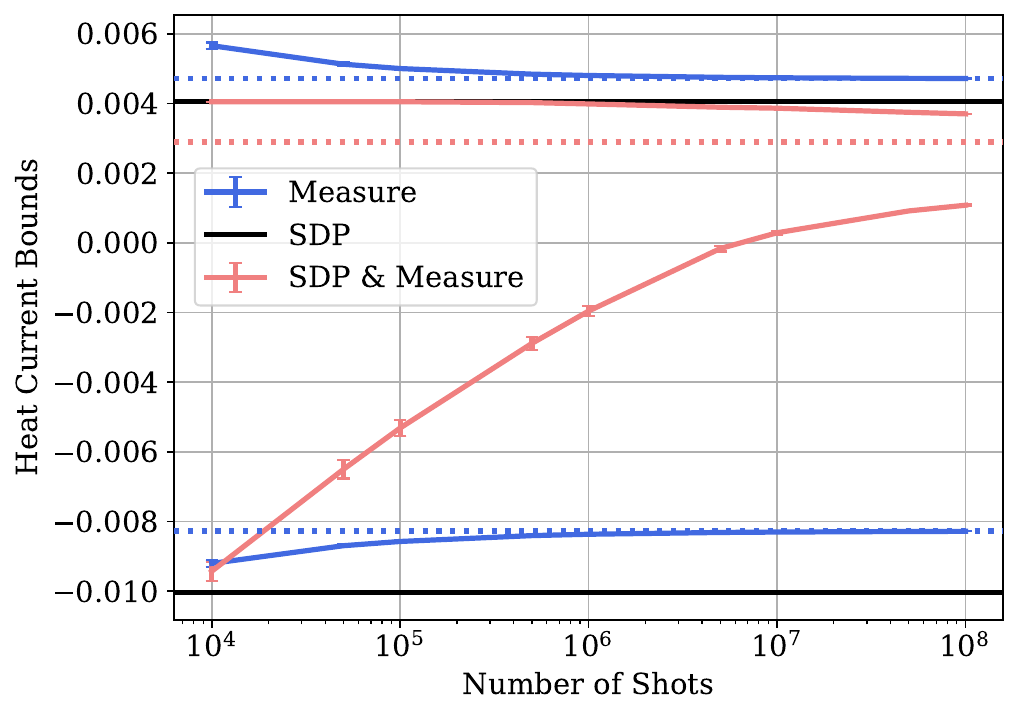}
\caption{Plot showing how the bounds on the heat current change versus the total number of shots $N_{\rm tot}$. 
Each strategy is identified by a different colour, and for a given strategy the upper curve represents the upper bound, while the lower curve represents the lower bound.
All measurements strategies here are without Z measurements. ``Measure'' refers to measuring the first 100 Pauli strings that appear in the successive generation of Lindbladian constraints, ``SDP'' refers to the SDP relaxation {without measurements (formally, Eq.\,\eqref{sdp-notation} taking $K=0$)}, and ``SDP \& Measure'' refers to the combination of both. 
The dashed lines represent the bound in the limit of infinite shots. 
Bounds involving measurements are valid under a confidence level of $99.7\%$. 
Each point is from $50$ repeats. Our SDP here uses a moment matrix of size $150\times150$, as well as $2000$ linear constraints from the Lindbladian. The open system here is a $3\times3$ grid of qubits with Lindbladian given by Eq.\,\eqref{eqn:lindblad}, with parameters $\gamma_c=0.011$, $\gamma_h=0.001$, $T_h=1.0$, $T_c=0.1$, and $J=h=1$. 
{Notably, the combined strategy (SDP \& Measure) is the only one able to certify the positivity of the heat current, as the lower bound is greater than zero. Overall, its results are substantially more informative than the ones obtained via the other two strategies.}}
\label{fig:heat}
\end{figure}

\subsection{Ground-state Energy}
We now turn to the case of obtaining a lower bound on the ground-state energy. 
We again consider the Hamiltonian in Eq.~\eqref{eqn:hamil}, corresponding to a two-dimensional transverse-field Ising model.
Hence, the objective function is given by the following linear combination of moment variables:
\begin{equation}
\label{ham-obj}
\langle \mathcal{O}\rangle\equiv 
g\sum_{i}^n \langle Z_i \rangle + \frac{J}{2}  \sum_{\braket{i,j}}^n \langle X_i X_j \rangle \,.
\end{equation}
In Fig.~\ref{fig:energy}, we plot the lower bound on the ground-state energy as a function of the total number of simulated measurement shots.
Although the SDP without measurements already yields a reasonably tight lower bound, we observe that the bound is further improved even with a relatively small number of shots.
More specifically, in Fig.~\ref{fig:energy} we present results for different variants of our method, corresponding to different choices of variables and constraints.
The best sampling method for few-shots appears to be measuring the most common Pauli strings, in this case the $100$ most common. 
Meanwhile, measuring only the Pauli strings that appear in the objective function \eqref{ham-obj} initially gives worse results, but eventually beats the automatic selection once sufficient measurements are performed.
Measuring all second-order Pauli strings, instead, typically yields performance intermediate between the two methods.

\begin{figure}
    \centering
    \includegraphics[width=0.99\linewidth]{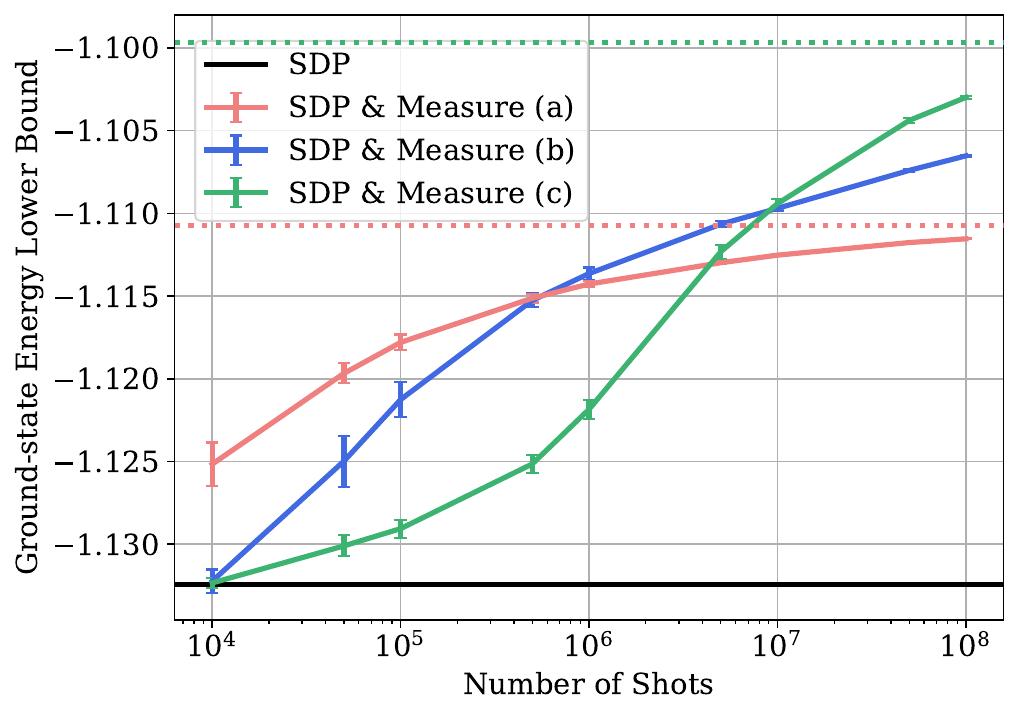}
    \caption{Plot showing how the lower bound on the ground-state energy changes versus the total number of shots. Here ``SDP \& Measure (a)'' refers to measuring the 100 Pauli strings that appear most frequently in the SDP, ``SDP \& Measure (b)'' refers to measuring all second-order Pauli strings, ``SDP \& Measure (c)'' refers to measuring only the Pauli strings that appear in the objective, whilst ``SDP'' refers to the SDP relaxation of the problem without any measurement data. The dashed lines represent the bound in the limit of infinite shots. Sets (b) and (c) have the same bound at infinite shots - the true ground state. All results here are to a confidence level of $99.7\%$. Each point is the mean from 50 repeats. Our SDP here uses a moment matrix of size $150\times150$. The system here is a $3\times3$ grid of qubits with Hamiltonian given by Eq.\,\eqref{eqn:hamil} for $J=h=1$.}
    \label{fig:energy}
\end{figure}

{\color{violet}
{\color{black}
\subsection{Entanglement and purity}
The purity of a state is given by $\mathcal{P}(\rho)=\tr{\rho^2}$, which in terms of Pauli strings can be written as a nonlinear function of the moments $({1}/{2^n}) \sum_\alpha \braket{P_\alpha}^2$.
To further demonstrate the versatility of our method, we move now to the calculation of lower bounds of purity or, equivalently speaking, upper bounds for the linear entropy
$S_{\mathrm{lin}}(\rho) \equiv 1 - {\rm tr}(\rho^2)$.

Interestingly, purity can be used as a witness of entanglement under certain 
conditions.
Entanglement in a pure bipartite state $\rho_{AB}$ can be revealed by comparing the entropy of a subsystem with that of the whole system. In particular, when the reduced state $\rho_A$ has larger entropy than the global state, this imbalance certifies the presence of bipartite entanglement across the $A|B$ cut. Thus, the entropy of a local subsystem provides a natural diagnostic of how strongly $A$ is correlated with the rest of the system \cite{horodecki2009quantum, brydges2019probing}. 
For our purposes it is useful to explicitly write the linear entropy of the reduced state $S_{\rm lin}(\rho_A)=1-\tr{\rho_A^2}$--in essence, the entanglement quantifier--%
in terms of the moments of A,
namely
\begin{eqnarray}
\label{mixednessA}
S_{\rm lin}(\rho_A)&=&1-\frac{1}{d_A}
\sum_{P_A} \langle P_A \rangle^{2}\,,
\label{lin-entropy}
\end{eqnarray}
and recognize that this is a concave function in the moments. 
The latter property allows one to maximize this function within polynomial optimization frameworks.
However, we have to guarantee scalability of the approach.
On this regard, we can choose $d_A^2=4^{n_A}$ to be $poly(n)$; for instance merely keeping fixed (i.e., independent of $n$) the number $n_A$ of qubits that we assign to $A$. This ensures our problem remains scalable in $n$.

Instead, if we want to bound the \enquote{global} purity of the $n$-qubit state 
we have to find a caveat as it involves
$O(4^n)$ variables.
On this regard, we notice that by truncating the sum 
to a scalable set of variables (identified by the index set $\mathcal{I}_o$, see Table\,\ref{tab:sets})
we obtain a lower bound on it, i.e.,
\begin{equation}
\label{truncated-purity}
    \tilde{\mathcal{P}}:=\frac{1}{d}
\sum_{\gamma \in \mathcal{I}_o} \langle P_\gamma \rangle^2 \leq 
\mathcal{P} \,.
\end{equation}
Minimizing 
$\tilde{\mathcal{P}}$
is now a manageable task within our approach. 

In the example that we shall choose, the true state $\rho$ is the non-degenerate ground state of a given Hamiltonian.
It is therefore a pure state, i.e. the true value of the purity is $\mathcal{P}(\rho)=1$.
To force ground-state properties,
as in Ref.\,\cite{PhysRevX.14.031006}, we notice that one could constrain the system energy to lie within an energy shell that contains the ground-state energy, 
\begin{equation}
\label{energy-shell}
E_{\rm gs}^{\rm (lb)} \leq \langle H \rangle 
\leq E_{\rm gs}^{\rm (ub)}\,.
\end{equation}
Interestingly, while in Ref.\,\cite{PhysRevX.14.031006}
the lower bound should be obtained by an appropriate SDP relaxation 
and the upper bound by variational methods, 
within our approach an energy window can be retrieved from measurement data, by including the Hamiltonian itself among the measured observables $\{M_k\}_k$.
Similarly to Ref.\,\cite{PhysRevX.14.031006}, one can use the constraint \eqref{energy-shell} as the basis for bounding additional system properties.
As in the previous examples, this can be further strengthened by incorporating information obtained from other measurement data.

Taken together, these considerations underscore the broad applicability of our measurement-enhanced approach.
Concretely, in the numerical simulations that we show, we focus on the ground state properties of the 2D transverse-field Ising Hamiltonian in Eq.\,\eqref{eqn:hamil}.
}%
{%
%
}%
}%
%
{\color{black}%
We compute the 
{global purity lower bound} minimizing $\tilde{\mathcal{P}}$ in Eq.\,\eqref{truncated-purity}, serving as our objective function.
Our results, shown in Figure \ref{fig:purity}, demonstrate that combining measurement results with an SDP approach allows, once again, for a better bound. In this case, neither method reaches the true optimum of $1$ even at the limit of infinite shots, as we only consider up to second-order Pauli strings rather than the ninth-order (meaning $4^9=262144$) that would be needed for this $3\times 3$ system. As such, we believe it is remarkable to reach a lower bound of $10\%$ of the optimum using only $0.1\%$ of the possible Pauli strings.
}
}

\begin{figure}[t!]
    \centering
    \includegraphics[width=0.99\linewidth]{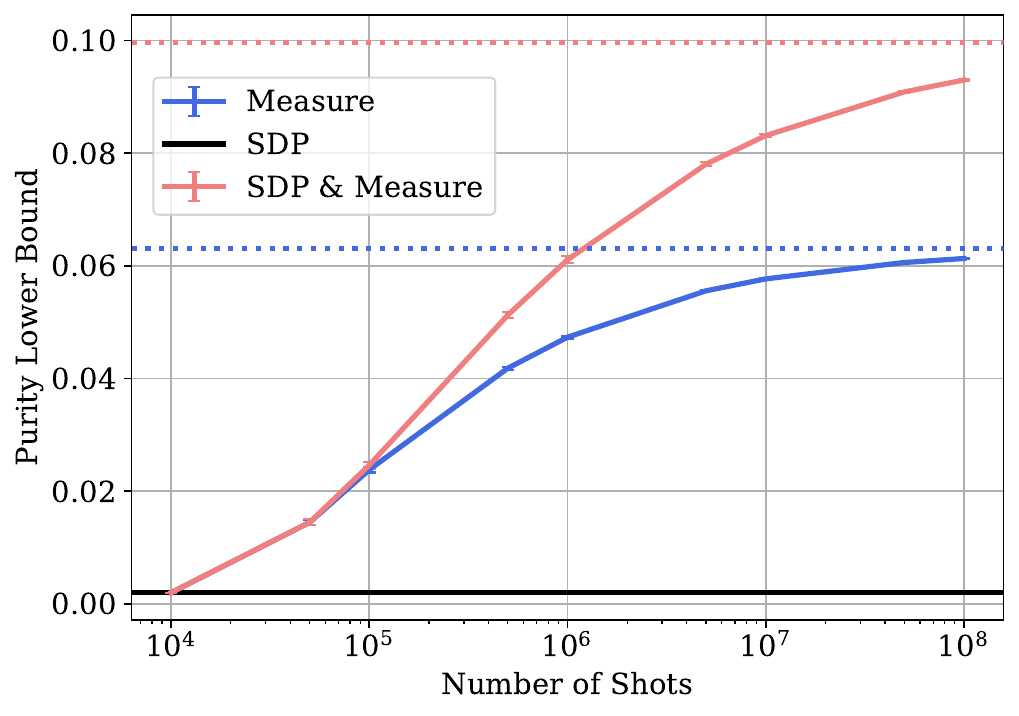}
    \caption{Plot showing how the lower bound on the purity of the ground state changes versus the total number of shots. The known optimum is $1$ as the ground state is pure. Here ``Measure'' refers to measuring all second-order Pauli strings, ``SDP'' refers to the SDP relaxation {without measurements}, and ``SDP \& Measure'' refers to the combination of both. The dashed lines represent the bound in the limit of infinite shots. All results here are to a confidence level of $99.7\%$. Each point is the mean from 50 repeats. Our SDP here uses a moment matrix of size $150\times150$. The system is a $3\times3$ grid of qubits with Hamiltonian given by Eq.\,\eqref{eqn:hamil} for $J=h=1$.}
    \label{fig:purity}
\end{figure}

\subsection{Larger Systems}
\label{sec:large-system}
To see how the number of shots might affect the SDP performance for larger systems, we use our method to find a lower bound for the ground state energy of the well-studied Majumdar–Ghosh model \cite{majumdar1969next} for 50 qubits. This system has a known ground-state, and as such it allows us to sample efficiently from the true expectation values of the Pauli strings. The model has a Hamiltonian of:
\begin{equation}
    \label{eqn:mg}
\begin{aligned}
     {\hat {H}} &= \sum _{j=1}^{N}({X}_{j}{X}_{j+1}+{Y}_{j}{Y}_{j+1}+{Z}_{j}{Z}_{j+1}) \\
     &+ {\frac {1}{2}}\sum _{j=1}^{N}({X}_{j}{X}_{j+2}+{Y}_{j}{Y}_{j+2}+{Z}_{j}{Z}_{j+2})
\end{aligned}
\end{equation}
and a known ground state such that all Pauli strings have a value of zero if the Pauli string is not formed of pairs acting on adjacent qubits (e.g. $\braket{X_1X_2}\ne0$, $\braket{X_1X_3}=0$). For the non-zero Pauli strings, the value is $(-1)^{m/2}$ where $m$ is the number of nontrivial Pauli operators in the string (i.e., different from the identity).

Using this ability to efficiently simulate samples even for large system sizes, we demonstrate how combining the SDP relaxation and such measurement data,  one obtains tighter lower bounds on the the ground state energy
than either method individually. 
The known ground-state energy per qubit is $-0.375$, exactly the value that we converge to in the limit of infinite shots. These results are given in Figure \ref{fig:large}.

\begin{figure}
    \centering
    \includegraphics[width=0.99\linewidth]{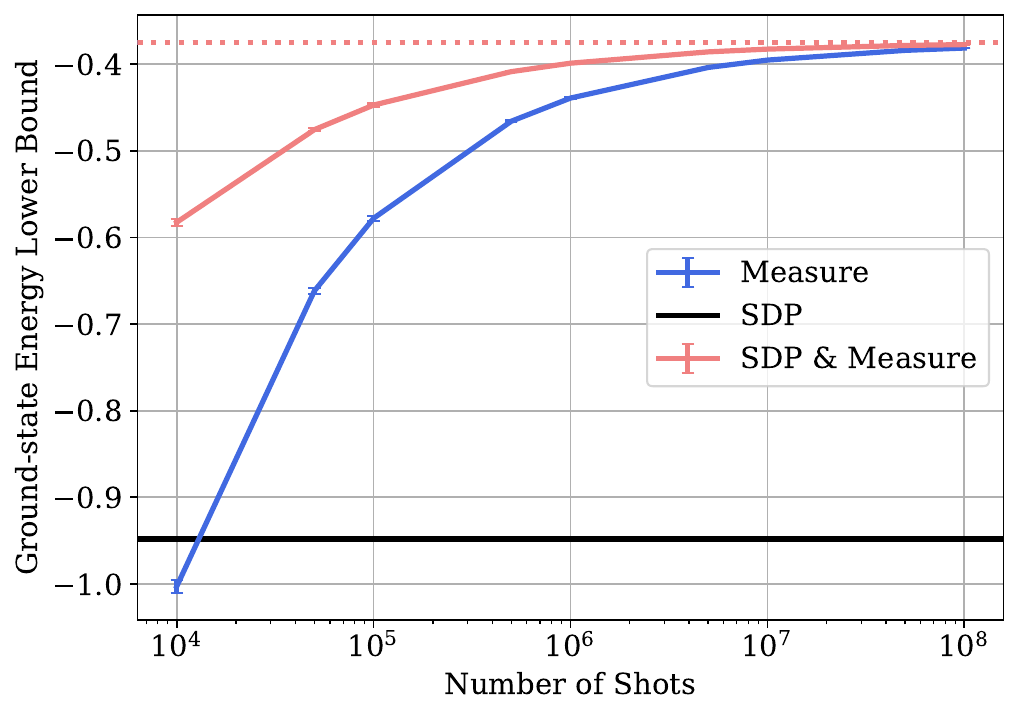}
    \caption{Plot showing how the lower bound on the ground-state energy for the 50-qubit system changes versus the total number of shots. Here ``Measure'' refers to measuring the Pauli strings that appear in the objective, ``SDP'' refers to the SDP relaxation, and ``SDP \& Measure'' refers to the combination of both. The dashed lines represent the bound in the limit of infinite shots. All results here are to a confidence level of $99.7\%$. Each point is the mean from 50 repeats. Our SDP here uses a moment matrix of size $200\times200$. The system is a 50-qubit Majumdar–Ghosh model, the Hamiltonian given as Eq.\,\eqref{eqn:mg}.}
    \label{fig:large}
\end{figure}

\subsection{Confidence Regions}
\label{sec:conf-region-plot}
A key point separating our method from other approaches is the ability to generate bounds with different confidence levels. Whilst for all the results shown 
we used 99.7\% as confidence level, we demonstrate in Fig.\,\ref{fig:confidence} how the purity lower bound would look using 68\%, 95\% and 99.7\%. By changing this variable, we change the $\epsilon_i(\delta)$'s from Eq.\,\eqref{eqn:epsilon} and thus the bounds on the Pauli strings in our SDP relaxation, see problem \,\eqref{sdp-notation}. 
The results are as expected: considering the problem of calculating a lower bound, if one requires a stronger assurance $1-\delta$ that the bound is valid, the bound must be lowered, i.e. less tight.
However, as $1/\delta$ appears only within the logarithm in Eq.~\eqref{eqn:epsilon},
increasing the confidence level generally results in only slightly looser bounds.

\begin{figure}[t]
    \centering
    \includegraphics[width=0.99\linewidth]{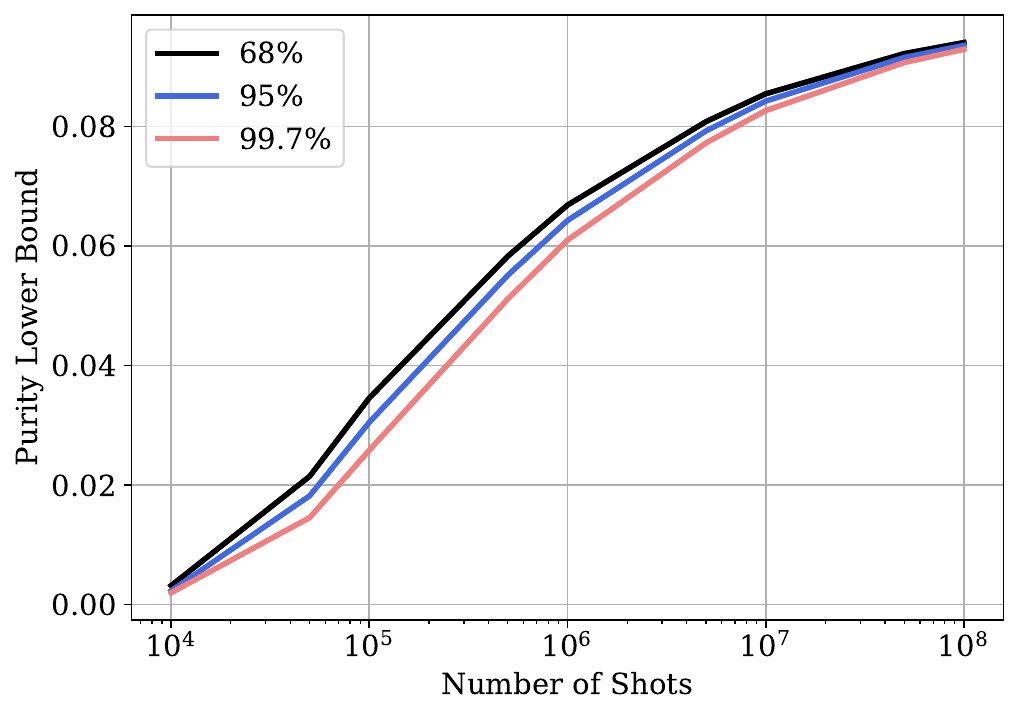}
    \caption{Plot showing how the lower bound on the purity changes versus both the total number of shots (abscissa) and versus the desired confidence level $1-\delta$ (legend).
    {Increasing the confidence level results in only slightly looser bounds, as $1/\delta$ appears only within the logarithm in Eq.~\eqref{eqn:epsilon}.}
    In this case each line is obtained using the SDP relaxation combined with measurement data of all second-order Pauli strings. Our SDP here uses a moment matrix of size $150\times150$. 
    The system here is a $3\times3$ grid of qubits with Hamiltonian given by Equation \eqref{eqn:hamil} for $J=h=1$.}
    \label{fig:confidence}
\end{figure}

\section{Discussion}
\label{sec:conc}

In this work we combined the methods of semidefinite programming relaxations and tomography to obtain probabilistic bounds on arbitrary observables for many-body quantum systems. This allows one to obtain tighter bounds than those of either method alone. 
We first introduced the approach in full generality and then applied it to paradigmatic problems in many-body quantum information.
We demonstrated an advantage in tightening the range of possible values for the heat current characterizing the nonequilibrium steady state of an open quantum system.
Analogous clear advantages are observed concerning ground state properties, such as ground state energy and associated linear entropies.
%
%
%
%
Our method is completely versatile and capable of generating bounds with different confidence levels, providing an advanced tool for studying quantum systems when one has both measurement data and additional guarantees on the system state.

{From a computational standpoint, our results illustrate that moment-matrix relaxations provide a powerful and scalable alternative to full state tomography. While the bounds obtained are, by construction, less tight than those achievable with full information, the polynomial scaling makes the approach applicable to systems of tens or even hundreds of qubits. This scalability could be further enhanced by exploiting hierarchical relaxations that adaptively increase moment-matrix order only where necessary, or by using symmetries to reduce redundancy in the constraints.}

Our work shares a conceptual lineage with classical shadow tomography, which also aims to estimate properties from randomized measurements \cite{huang2020predicting}. However, standard shadow tomography typically does not leverage prior physical knowledge, such as the system being in a ground state or a steady state. By enforcing these constraints via the SDP, our method effectively restricts the search space to the physically relevant manifold.

Future work would include further testing and refinement of the scheme, applying it to other quantum systems and taking advantage of intrinsic symmetries. 
In particular, our approach could be used to derive upper bounds on the linear stabilizer  entropy \cite{PhysRevLett.128.050402, PhysRevA.110.L040403}, providing a tool to study the problem of maximal magic, alongside verifying entanglement via negativity bounds \cite{Moroder2013}.
The true test would be to then use real experimental data to explore systems that one cannot easily simulate.
We foresee useful applications of our approach across experimental platforms that engineer spin-chain models, such as Rydberg-atom arrays and trapped-ion setups.
This would provide a confidence-certified framework for many-body quantum information, enabling reliable certification of nonclassical behavior in quantum technologies.

\begin{acknowledgments}

This project has received funding from the European Union’s Horizon 2020 research and innovation programme under the Marie Skłodowska-Curie grant agreement No 847517, 
PNRR MUR Project No. PE0000023-NQSTI,
University of Catania via PNRR-MUR Starting Grant project PE0000023-NQSTI,
the Government of Spain (Severo Ochoa CEX2019-000910-S, FUNQIP and European Union NextGenerationEU PRTR-C17.I1), Fundació Cellex, Fundació Mir-Puig, Generalitat de Catalunya (CERCA program), the EU Quantera project Veriqtas and COMPUTE.

\end{acknowledgments}

\bibliography{bibfile}

\appendix

\begin{widetext}

\section{Further details on positivity relaxations}
\label{app:pos}
For completeness we illustrate how moment matrices and reduced density matrices take the generic form
\begin{equation}
    B_k+\textstyle{\sum_{\alpha \in \mathcal{I}_{+_k}}} A_{k,\alpha} \braket{P_{\alpha}}\,, 
\end{equation}
reported in problem\,\eqref{sdp-notation}, $k$ labelling a chosen relaxation.
We show this through minimal examples from which 
it is evident how the general statement holds true.
Let us consider the moment matrix
$(\mathcal{M}_{k})_{\alpha, \beta}=\tr{P_\alpha P_\beta \rho}$, for the vector 
$\{P_\alpha\}_\alpha:=(\mathbb{1},Z_1,Z_2,X_1X_2)^T$.
Such moment matrix takes the form
(using Pauli reduction rules),
\begin{equation}
    \begin{pmatrix}
        1& \langle Z_1\rangle  &\langle  Z_2\rangle  &\langle  X_1 X_2\rangle \\
       \langle Z_1 \rangle & 1& \langle Z_1 Z_2 \rangle  & i \langle Y_1 X_2\rangle  \\
        \langle Z_2 \rangle & \langle Z_1 Z_2 \rangle & 1& i \langle X_1 Y_2 \rangle \\
        \langle X_1 X_2 \rangle & -i \langle Y_1 X_2 \rangle & -i \langle X_1 Y_2 \rangle & 1
    \end{pmatrix}
\end{equation}
and can be cast as the sum 
\begin{equation}
B+A_1 \langle Z_1 \rangle
+A_1 \langle Z_1 \rangle
+A_2 \langle Z_2 \rangle
+\dots +
A_6 \langle X_1 Y_2 \rangle\,,
\end{equation}
with 
\begin{equation}
B=\mathbb{1}_4,\,
A_1=\begin{pmatrix}
0& 1& 0& 0 \\
1& 0& 0& 0 \\
0& 0& 0& 0 \\
0& 0& 0& 0 
\end{pmatrix}
,\,
A_2=\begin{pmatrix}
0& 0& 1& 0 \\
0& 0& 0& 0 \\
1& 0& 0& 0 \\
0& 0& 0& 0 
\end{pmatrix}
, \dots,\,
A_6=\begin{pmatrix}
0& 0& 0& 0 \\
0& 0& 0& 0 \\
0& 0& 0&i& \\
0& 0&-i& 0& 
\end{pmatrix}\,.
\end{equation}
Moment-matrix positivity is a relaxation of state positivity. To see this, we have to show that  
\begin{equation}
    \rho \succcurlyeq 0 \implies \mathcal{M}_k \succcurlyeq 0\,. 
\end{equation}
By definition 
$\mathcal{M}_k \succcurlyeq 0$
if and only if 
$\sum_{\alpha, \beta} v_\alpha^* (\mathcal{M}_k)_{\alpha \beta} v_\beta\geq 0$ for any complex vector $\{v_\alpha\}_\alpha$. 
We notice that
$\sum_{\alpha, \beta} v_\alpha^* (\mathcal{M}_k)_{\alpha \beta} v_\beta= {\rm tr}(\theta^\dagger \theta \rho)$, with $\theta:=\sum_\alpha v_\alpha P_\alpha $. 
A sufficient condition for the latter to be non-negative is $\rho$ being a positive operator (for a comprehensive discussion of the NPA hierarchy, we refer the reader to the original papers \cite{Navascues2007NPA,Navascues2008NPA}).
This holds true for arbitrary choices of operators used to build the moment matrix, demonstrating the thesis.

Concerning
reduced density matrices, for simplicity 
let us consider the one for the first two qubits.
It
takes already the desired form when written in the Pauli basis,
\begin{equation}
\rho_{12}=
\frac{1}{4}
\sum_{\alpha, \beta=0}^3 
\sigma_\alpha \otimes \sigma_\beta
\,
\langle \sigma_\alpha \otimes \sigma_\beta\rangle \,,
\end{equation}
where we have used here the more convenient notation
$(\sigma_0,\sigma_1, \sigma_2, \sigma_3)
\equiv
(\mathbb{1}_2, X, Y, Z)$.

As for moment matrix, reduced density matrices can be used 
to impose relaxations of state positivity because
\begin{equation}
    \rho \succcurlyeq 0 \implies \rho_k \succcurlyeq 0\,, 
\end{equation}
$\rho_k$ being an arbitrary reduced density matrix.

\clearpage

\end{widetext}
\end{document}